\documentclass[twocolumn,showpacs,preprintnumbers,amsmath,amssymb]{revtex4}

\usepackage{graphicx}
\usepackage{dcolumn}
\usepackage{bm}

\usepackage{psfrag}
\usepackage[latin1]{inputenc}
\usepackage{amsmath}
\usepackage{amsfonts}
\usepackage{amssymb}
\usepackage{bbm}

\begin{document}

\title{Verification of the Rayleigh Scattering Cross Section}
\author{Sayan Chakraborti}
\affiliation{Tata Institute of Fundamental Research, Mumbai, India}
\email{sayan@tifr.res.in}

\date{\today}


\begin{abstract}
A simple experiment is described for the direct determination of the wavelength dependence of the Rayleigh scattering cross section using the classical example of the blue sky.
\end{abstract}

\maketitle

\section{Introduction}
Rayleigh Scattering \cite{Rayleigh} has always been invoked to explain the blueness of the sky. It is the process of scattering of light by particles much smaller than the wavelength of the incident electromagnetic radiation. In this simple experiment the wavelength dependence of the Rayleigh scattering
cross section is determined using a commercially available spectrometer.

\section{Theory}
Following Jackson \cite{Jackson} we shall treat the scattering of radiation of frequency $\omega$ by a single nonrelativistic particle of mass $m$ and charge $e$ under a spherically symmetric, linear restoring force $m \omega_0^2 x$. The equation of motion is written as
\begin{equation}
m(\mathbf{\ddot x} - \tau \mathbf{\dddot x} + \omega_0^2 \mathbf{x})=\mathbf{F}(t)
\end{equation}
where the characteristic time is given by
\begin{equation}
\tau = \frac{2}{3} \frac{e^2}{m c^3}
\end{equation}
Adding a resistive term to the left hand side, our equation becomes
\begin{equation}
\mathbf{\ddot x} + \Gamma' \mathbf{\dot x} - \tau \mathbf{\dddot x} + \omega_0^2 \mathbf{x} = \frac{e}{m} \mathbf{\epsilon} E_0 e^{-i \omega t} 
\end{equation}
where $E_0$ is the electric field and $\mathbf{\epsilon}$ is the incident polarization vector. The steady state solution is given by
\begin{equation}
\mathbf{x}=\frac{e}{m} \frac{E_0 e^{-i \omega t}}{\omega_0^2 - \omega^2 - i \omega \Gamma_t} \mathbf{\epsilon}
\end{equation}
where we have written the total decay constant or total width as
\begin{equation}
\Gamma_t(\omega) = \Gamma' + \frac{\omega}{\omega_0}^2 \Gamma
\end{equation}
The radiative decay constant is $\Gamma=\omega_0^2\tau$. The radiation field caused by this accelerated motion is thus given by
\begin{equation}
\mathbf{E}_{rad}=\frac{e}{c^2} \frac{1}{r}[\mathbf{n \times (n \times \ddot x)}]_{ret}
\end{equation}

Hence the radiation field in any particular direction denoted by the polarization $\mathbf{\epsilon'}$ is given by
\begin{equation}
\mathbf{\epsilon'.E}_{rad}=\frac{e^2}{mc^2} \omega^2 \frac{E_0 e^{-i \omega t} e^{ikr}}{\omega_0^2 - \omega^2 - i \omega \Gamma_t}\Big(\frac{\mathbf{\epsilon.\epsilon'}}{r}\Big)
\end{equation}
From the definition of differential cross section the cross section for scattered light of frequency $\omega$ and polarization $\mathbf{\epsilon'}$ is written as
\begin{eqnarray}
\frac{d \sigma (\omega,\mathbf{\epsilon'})}{d \Omega} = \Big\vert\frac{r\mathbf{\epsilon'.E}_{rad}}{E_0}\Big\vert^2 \nonumber\\
= \Big(\frac{e^2}{mc^2}\Big)^2 (\mathbf{\epsilon.\epsilon'})^2 \Big[\frac{\omega^4}{(\omega_0^2-\omega^2)^2+\omega^2\Gamma_t^2}\Big]
\end{eqnarray}
In the large wavelength limit, the scattering cross section reduces to
\begin{equation}
\frac{d \sigma (\omega,\mathbf{\epsilon'})}{d \Omega} = \Big(\frac{e^2}{mc^2}\Big)^2 (\mathbf{\epsilon.\epsilon'})^2 \Big(\frac{\omega}{\omega_0}\Big)^4
\end{equation}
This gives a $\lambda^{-4}$ dependence to the scattering cross section at large wavelengths.

\section{Experiment}

\begin{figure*}
\input{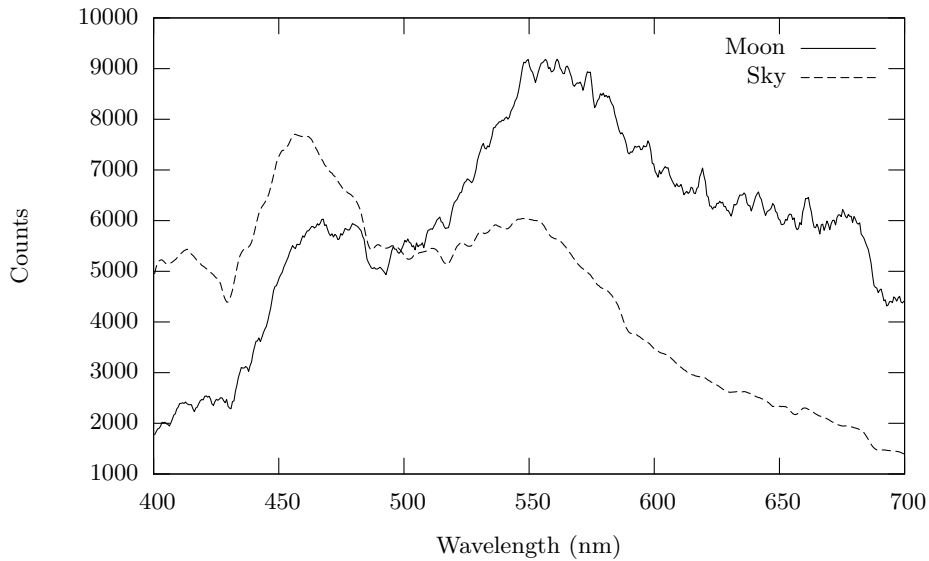}
\caption{\label{fig:1} Spectra of moon and sky}
\end{figure*}

The idea is to compare the spectrum of a blue sky with that of the solar spectrum to find out the
scattering cross section at each wavelength. However a direct solar spectra could not be taken with the instrument at hand, as it was sataurated even at the lowest allowed exposure of 2ms. The lunar spectra which is very similar to the solar spectra was used instead.

An AvaSpec-2048 Fiber Optic Spectrometer was used to record the respective spectra and averaged over 10 readings and suitably dark subtracted. It was ensured that all counts in the region of interest, namely 400 to 700 nm, were above 1000. Since the data was averaged over 10 cycles, it represents a photon count of more than 10000. Assuming Poisson noise, the fractional error at each wavelength is less than 1 percent.

Evidently the sky spectra is bluer than the lunar spectra. This blueness of the sky is supposedly due to Rayleigh scattering by the atmosphere. We shall now try to extract its wavelength dependence and see if it matches with theory.

\section{Data Reduction}

\begin{figure*}
\input{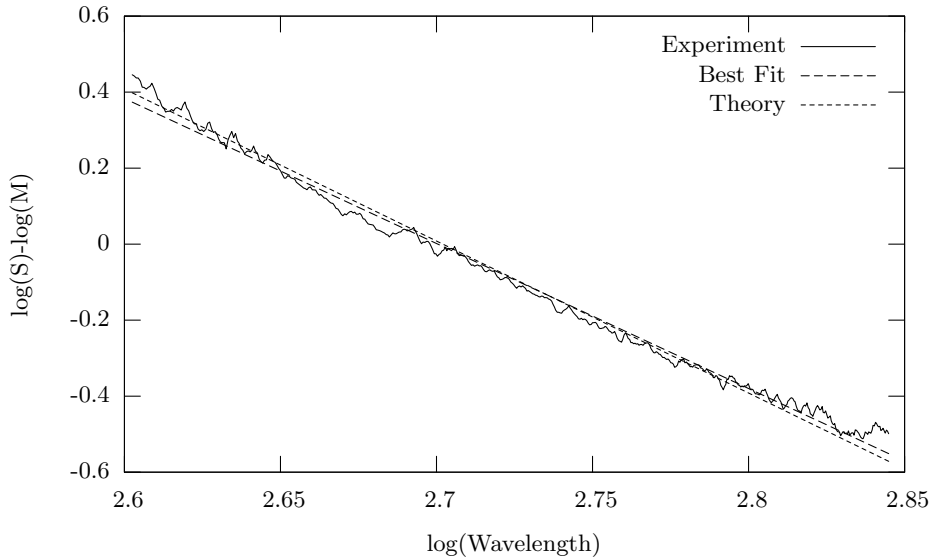}
\caption{\label{fig:2} Relative scattering cross section}
\end{figure*}

Before any can conclusions can be drawn from the graphs, care must be taken to analyze what each count represents. The counts in the lunar spectra can be represented as
\begin{eqnarray}
M(\lambda)=Constants \times Efficiency (\lambda)\nonumber\\
\times Albedo (\lambda)\times Solar Spectra(\lambda)
\end{eqnarray}
The counts in the sky spectra are on the other hand
\begin{eqnarray}
S(\lambda)=Constants \times Efficiency (\lambda)\nonumber\\
\times Rayleigh (\lambda)\times Solar Spectra(\lambda)
\end{eqnarray}
We wish find out the wavelength dependence of the Rayleigh scattering cross section from this data. The quantum efficiency of the detector and the solar spectra can be eliminated by the simple process of taking a ratio of the two spectra. The lunar albedo is approximately constant over the wavelength region being studied. As the Moon lacks an atmosphere, the average particle size of the scatterers on the lunar surface is much larger than incident wavelength.

We take logarithms to the base 10 of both spectra and plot their difference against the logarithm of the wavelength. The best fit straight line is found to have a slope of $-3.8$. The theoretical curve with a slope of $-4.0$ is also plotted for comparison.

\section{Conclusions}
The wavelength dependence of the Rayleigh scattering cross section is seen to approximately match the $\lambda^{-4}$ nature predicted by theory. The experiment is suitable for undergraduates and can be completed by taking a lunar spectra one night and the sky spectra on the next morning. It provides a direct way of verifying the $\lambda^{-4}$ nature of the Rayleigh scattering cross section which is taught in almost all undergraduate curricula.

\section{Acknowledgements}
The author would like to thank Prof. D.K. Ojha and Prof A.K. Ray for their guidance, Tarak Thakore and Rajesh for help with observations and Mehuli Mondal for discussions on Rayleigh scattering.

\end{document}